\documentclass[prb,twocolumn,amsmath,amssymb,superscriptaddress,floatfix]{revtex4}

\pdfoutput=1

\usepackage{amsmath}
\usepackage{amsfonts}
\usepackage{float}
\usepackage{ graphicx } 
\usepackage{ bm } 
\usepackage{ color }
\usepackage{ epstopdf }
\usepackage{enumerate}

\newcommand{\be}{\begin{equation}}
\newcommand{\ee}{\end{equation}}
\newcommand{\bea}{\begin{eqnarray}}
\newcommand{\eea}{\end{eqnarray}}

\newcommand{\ket}[1]{\left | #1 \right\rangle}
\newcommand{\bra}[1]{\left\langle #1 \right|}

\begin{document}
\title{Detection of topological phases by quasi-local operators}

\affiliation{Centre for Quantum Technologies, National University of Singapore, 3 Science Drive 2, Singapore 117543, Singapore}
\affiliation{ \textit CeFEMA,
Instituto Superior T\'ecnico, Universidade de Lisboa, Av. Rovisco Pais, 1049-001 Lisboa, Portugal}
\affiliation{\textit Beijing Computational Science Research Center, Beijing, China}
\affiliation{ \textit College of Materials Science and
Opto-Electronic Technology, University of Chinese Academy of
Sciences, Beijing, China }
\affiliation{School of Electrical and Computer Engineering, Technical University of Crete, Chania, Crete, 73100 Greece}

\author{ Wing Chi Yu }
\affiliation{Centre for Quantum Technologies, National University of Singapore, 3 Science Drive 2, Singapore 117543, Singapore}
\author{ P. D. Sacramento }
\email{pdss@cefema.tecnico.ulisboa.pt}
\affiliation{ \textit CeFEMA,
Instituto Superior T\'ecnico, Universidade de Lisboa, Av. Rovisco Pais, 1049-001 Lisboa, Portugal}
\affiliation{\textit Beijing Computational Science Research Center, Beijing, China}
\author{ Yan Chao Li }
\affiliation{ \textit College of Materials Science and
Opto-Electronic Technology, University of Chinese Academy of
Sciences, Beijing, China }
\affiliation{\textit Beijing Computational Science Research Center, Beijing, China}
\author{D.G. Angelakis}
\affiliation{Centre for Quantum Technologies, National University of Singapore, 3 Science Drive 2, Singapore 117543, Singapore}
\affiliation{School of Electrical and Computer Engineering, Technical University of Crete, Chania, Crete, 73100 Greece}
\author{ Hai-Qing Lin }
\affiliation{\textit Beijing Computational Science Research Center, Beijing, China}

\date{ \today }


\begin{abstract}
It has been proposed recently by some of the authors that the quantum phase transition of
a topological insulator like the SSH model may be detected by the eigenvalues and eigenvectors of
the reduced density matrix. 
Here we further extend the scheme of identifying the order parameters by considering the SSH model with the addition of triplet superconductivity. This
model has a rich phase diagram due to the competition of the SSH "order" and the Kitaev "order",
which requires the introduction of four order parameters to describe the various topological
phases. We show how these order parameters can be expressed simply as averages of projection operators on the ground state at certain points deep in each phase and how one can simply obtain the phase boundaries.
 A scaling analysis in the vicinity of the transition lines is consistent with the quantum Ising universality class.
\end{abstract}

\maketitle

\section{Introduction}
In condensed matter physics, the order parameter plays an important role in the study of phase transitions. It characterizes the order of a phase and helps to detect the critical point. People usually rely on physical intuition or resort to methods such as group theory
and the renormalization group analysis to identify the order parameters of a many-body system. However, those methods require a prior knowledge in the symmetries of the Hamiltonian and do not always apply, especially to systems exhibiting topological phase transitions. This calls for a general and systematic scheme to derive the order parameters without the aid of such an empirical knowledge.

Recently, a proposal based on using the dominant eigenstates of the reduced density matrix of a many-body system has been established by some of the authors \cite{yu1,yu2}. Unlike the other schemes proposed \cite{Furukawa2006,Cheong2009,Henley2014}, the approach is non-variational. Our scheme has also been extended to the detection of the topological phase of topological insulators \cite{us,bermudez} such as the Su-Schrieffer-Heeger (SSH) model \cite{ssh},
or equivalently the Schockley model \cite{schockley}.

In our original proposal, one has to first determine the minimum size of the reduced density matrix that captures the extended correlations in the system by calculating the mutual information. Then the dominant states (with relatively larger eigenvalues) of the reduced density matrix are used to construct the
order parameter. For example, in the Mott insulator phase of the fermion Hubbard model, the single site reduced density matrix has largest eigenvalues for the local state spin up and spin down. One can then define the order operator as a linear combination of the projectors of these two states which turns out to be the Pauli matrix in the $z$ direction as expected \cite{yu2}. However, in some cases such as the topological phase of the SSH model, several eigenstates of the reduced density matrix with comparable weights contribute and lead to an undetermined combination in the order parameter \cite{us}.

In this work, we introduce a fundamental extension of our previous proposal \cite{us} to overcome the above mentioned issue by considering the projector to a subset of the system's ground state. The order parameter is then defined as the expectation value of this projector in the original representation. This method can be applied to any models that have a diagonal representation of the Hamiltonian at some specific points in the phase diagram. 
We apply the method to the SSH model with addition of triplet
pairings between the fermionic states. The phase
diagram of the model consists of a trivial phase, two topological phases of Kitaev type and one
topological phase of SSH-type. The four order parameters that describe the various phases of the model are obtained.

The paper is organized as follows. In Sec. \ref{sec:scheme} we recap our original scheme of deriving the order parameter and introduce the new method. We then apply the method to obtain the order operators in the SSH model with triplet pairing in Sec. \ref{sec:model} and calculate the order parameters (ground state average of the order operators) in Sec. \ref{sec:orderparameter}.  We also study the universality class of the model from finite size scaling analysis of the derived order parameters in Sec. \ref{sec:scaling}. In Sec. \ref{sec:localH}, we mention that the application of the method to the Kitaev model leads to a quasi-local operator that may be identified with the local Hamiltonian. It is shown
both for the Kitaev model and the SSH-Kitaev model that the local
Hamiltonian may also be used to detect the topological transitions. Finally, a conclusion is given in Sec. \ref{sec:conlusion}.

\section{The method}
\label{sec:scheme}
The first step in our original proposal \cite{yu1} to derive the order parameter is to determine the minimum size of the block (or subsystem) that the mutual
information does not vanish at a long distance.
The mutual information is defined as%
\begin{equation}
S(i,j)=S\left( {\rho }_{i}\right) +S\left( {\rho }_{j}\right) -S\left( {\rho
}_{i\cup j}\right) ,
\label{eq:MI}
\end{equation}%
where $S\left( {\rho }_{i}\right) =-\textrm{tr}({\rho }_{i}\ln{\rho
}_{i})$ is the von-Neumann entropy of the block $i$. The reduced density matrix $\rho_i$ is obtained by tracing out
all other degrees of freedom except those of the block $i$,
i.e. $\rho_{i}=\textrm{tr}\ket{\Psi_{0}}\bra{\Psi_{0}}$ where $\ket{\Psi_0}$ is the ground state of
the system. If and only if the mutual information is non-vanishing
at a long distance, there exists a long-range order (or quasi long range order) in the
system \cite{MMWolf,SJGuJPA}.

The next step is to calculate the eigenvalues and eigenstates of
the reduced density matrix of the desired block size. The order parameter is then defined as the linear combination of the dominant eigenstates \cite{yu1}, i.e.
\begin{eqnarray}
O_i=\sum_{\mu\leq\xi}w_{\mu}a_{i\mu}^{\dagger}a_{i\mu},
\label{eq:O_d}
\end{eqnarray}
where $a_{i\mu}^{\dagger} (a_{i\mu})$ is the creation (annihilation) operator of the state $\mu$ at site $i$, and $\xi$ is the rank of $\rho_i$. It can be proved that for any
$\mu>\xi$, the operator $a_{i\mu}^{\dagger}a_{i\mu}$ does not
correlate. The coefficients $w_{\mu}$ can be fixed by the traceless
condition tr$(\rho_iO_i)=0$ and the cut-off condition
$\max(\{w_{\mu}\})=1$.

In some cases, the basis of the reduced density matrix may not be ideal because of degeneracies and making it hard to determine the coefficients $\omega_{\mu}$ in Eq. \ref{eq:O_d}. In our previous work \cite{us} where the SSH model is considered, we demonstrated that such a difficulty can be overcome by a transformation into a Majorana basis. This allows a diagonal representation of the Hamiltonian in terms of fermionic operators, that are non-local combinations of the original fermion operators, at some specific points in the phase diagram. The dominant eigenstate of the reduced density matrix in this diagonal basis is then simply a subset of the system's groundstate.

This suggests one may try a different approach and motivates us to introduce a variation of our original scheme as used in this work. Consider a Hamiltonian which can be expressed in terms of some quasi-local Hamiltonians $H_j$. The quasi-local Hamiltonian is in general a function of a set of parameters, i.e. $H_j(g_1,g_2,\cdots)$. We may identify inside a phase in the phase diagram a point $(G_1,G_2,\cdots)$ where we can diagonalize the Hamiltonian. Call $|G_1,G_2,\cdots \rangle$ the groundstate of $H_j$ at
this point. Define an operator
\be
{\hat O}_j = |G_1,G_2,\cdots \rangle \langle G_1,G_2,\cdots |,
\ee
which is a projector to a subset of the system's groundstate.
We may now define the order parameter as the average value of this operator
in the groundstate of $H_j(g_1,g_2,\cdots)$.

Note that
\bea
O_j &=& \langle g_1,g_2,\cdots | {\hat O}_j |g_1,g_2,\cdots \rangle
\nonumber \\
&=& |\langle g_1,g_2,\cdots | G_1,G_2,\cdots \rangle |^2.
\eea
Physically, this is like a measure of the overlap between the ground state at two points in the phase diagram. However, unlike the conventional fidelity approach to quantum phase transitions \cite{Quan2006,Zanardi2006,GuFidelity}, the two points $(g_1,g_2\cdots)$ and $(G_1,G_2,\cdots)$ are in general far apart. See also for instance
\cite{tm,tm2}. In the next section, we illustrate the method in more detail by applying it to the SSH model with triplet pairing (SSH-Kitaev model).

\section{Order operators of SSH model with triplet pairing}
\label{sec:model}

\subsection{Model}

\begin{figure*}
\includegraphics[width=0.75\textwidth]{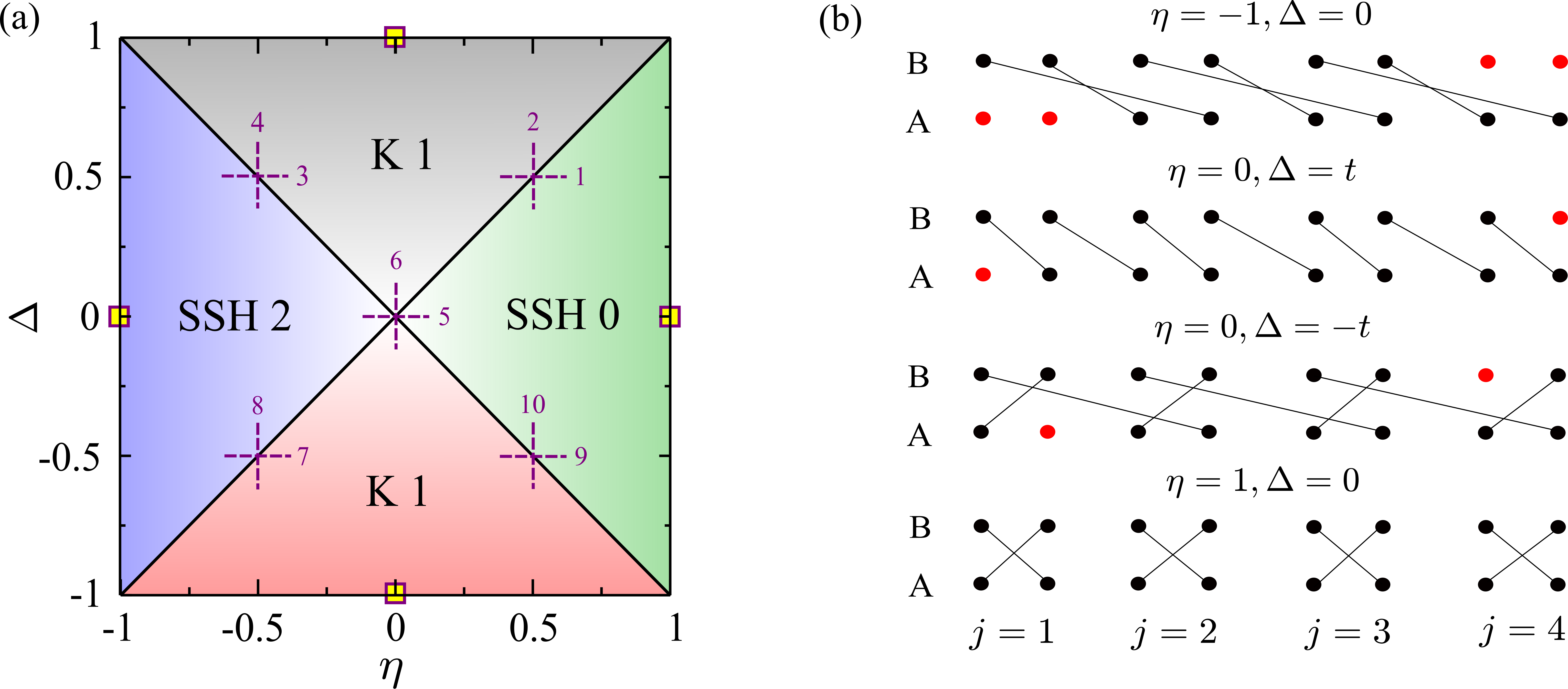}
\caption{\label{fig1}
(Color online)
(a) Phases of SSH-Kitaev model for zero chemical potential. When $\Delta=0$ the model reduces to the SSH model and for negative
$\eta$ the model is topologically non-trivial with edge states represented
by the decoupled Majorana operators. As in the Schockley model since at each
end site there are two decoupled Majoranas, these combine to form edge fermionic
modes. This constitutes phase SSH2 with $\eta=-1, \Delta=0$ and two edge modes.
If superconductivity is present, and there is no dimerization $\eta=0$, the model
reduces to the Kitaev model. The phase K1 with $\eta=0, \Delta=t$ has two
decoupled Majorana operators, one at each end, and therefore there is one Majorana
mode at each edge. The model interpolates between Majorana modes and fermionic
modes as the parameters change. There is also a trivial phase with no zero energy modes
denoted SSH0 which is similar to the trivial phase of the Schockley model.
The Hamiltonian in terms of Majoranas simplifies at the four points marked with
yellow squares and (b) shows an illustration of the Majorana modes at these four points respectively. At each of the lattice site $j$, the two dots represent the Majorana operators $\gamma_{j,\sigma,1}$ and $\gamma_{j,\sigma,2}$ ($\sigma=A$ or $B$). The lines represent the links between the Majorana operators.
}
\end{figure*}

This model may be viewed as a dimerized Kitaev superconductor \cite{tanaka}.
The dimerization is parametrized by $\eta$ and the superconductivity
by $\Delta$.

This model is given by the Hamiltonian
\bea
H = -\mu & \sum_j & \left(  c_{j,A}^{\dagger} c_{j,A} + c_{j,B}^{\dagger} c_{j,B} \right)
\nonumber \\
-t & \sum_j &  \left[  (1+\eta) c_{j,B}^{\dagger} c_{j,A} + (1+\eta) c_{j,A}^{\dagger} c_{j,B} \right.
\nonumber \\
&+& \left. (1-\eta) c_{j+1,A}^{\dagger} c_{j,B} +(1-\eta) c_{j,B}^{\dagger} c_{j+1,A} \right]
\nonumber \\
+\Delta  & \sum_j &  \left[  (1+\eta) c_{j,B}^{\dagger} c_{j,A}^{\dagger} + (1+\eta) c_{j,A} c_{j,B} \right.
\nonumber \\
&+& \left. (1-\eta) c_{j+1,A}^{\dagger} c_{j,B}^{\dagger} +(1-\eta) c_{j,B} c_{j+1,A} \right],
\nonumber \\
& &
\label{ham1}
\eea
where $t$ is the hopping, $\Delta$ is the pairing amplitude and $\mu$ is the chemical potential.
The model with no superconductivity ($\Delta=0$) is related to the Schockley model by
taking $t_1=t(1+\eta)$ and $t_2=t(1-\eta)$. The region of $\eta>0$ corresponds to $t_1>t_2$
and vice-versa for $\eta<0$.
The Hamiltonian in real space mixes nearest-neighbor sites and also has
local terms. We consider a system with $j=1,\cdots,N$ ($N$ sites A and $N$ sites B).
The local terms can be grouped in the matrix
\be
H_{j,j} = \left(\begin{array}{cccc}
-\mu & -t(1+\eta) & 0 & -\Delta (1+\eta) \\
-t(1+\eta) & -\mu  & \Delta (1+\eta) & 0  \\
0 & \Delta (1+\eta) & \mu & t(1+\eta) \\
-\Delta (1+\eta) & 0 & t(1+\eta) & \mu  \\
\end{array}\right).
\ee
The non-local terms to the nearest-neighbors can be written as
\be
H_{j,j+1} = \left(\begin{array}{cccc}
0 & 0 & 0 & 0 \\
-t(1-\eta) & 0  & -\Delta (1-\eta) & 0  \\
0 & 0 & 0 & 0 \\
\Delta (1-\eta) & 0 & t(1-\eta) & 0  \\
\end{array}\right),
\ee
and
\be
H_{j,j-1} = \left(\begin{array}{cccc}
0 & -t(1-\eta) & 0 & \Delta (1-\eta) \\
0 & 0  & 0 & 0  \\
0 & -\Delta (1-\eta) & 0 & t(1-\eta) \\
0 & 0 & 0 & 0  \\
\end{array}\right).
\ee
The Hamiltonian matrices are the matrix elements in a basis
\be
\left(\begin{array}{c}
c_{j,A} \\
c_{j,B} \\
c_{j,A}^{\dagger} \\
c_{j,B}^{\dagger} \\
 \end{array}\right)
\ee

In general, a  fermion operator may be written in terms of two hermitian operators, $\gamma_1, \gamma_2$, in the
following way
\bea
c_{j,\sigma} &=& \frac{1}{2} \left( \gamma_{j, \sigma, 1} + i \gamma_{j, \sigma, 2} \right) \nonumber \\
c_{j,\sigma}^{\dagger} &=& \frac{1}{2} \left( \gamma_{j, \sigma, 1} -
i \gamma_{j, \sigma, 2} \right)
\label{majos}
\eea
The index $\sigma$ represents internal degrees of freedom of the fermionic operator, such as spin
and/or sub-lattice index, and the $\gamma$ operators are hermitian and satisfy a Clifford algebra
\be
\{\gamma_m,\gamma_n \}=2 \delta_{nm} .
\ee

In terms of Majorana operators the Hamiltonian is written as
\bea
H &=& -\frac{\mu}{2} \sum_{j=1}^N \left( 2+i\gamma_{j,A,1} \gamma_{j,A,2}
+ i \gamma_{j,B,1} \gamma_{j,B,2} \right) \nonumber \\
&-& \frac{it}{2} (1+\eta) \sum_{j=1}^N \left(
\gamma_{j,B,1} \gamma_{j,A,2} + \gamma_{j,A,1} \gamma_{j,B,2} \right) \nonumber \\
&-& \frac{it}{2} (1-\eta) \sum_{j=1}^{N-1} \left(
\gamma_{j+1,A,1} \gamma_{j,B,2} + \gamma_{j,B,1} \gamma_{j+1,A,2} \right) \nonumber \\
&+& \frac{i\Delta}{2} (1+\eta) \sum_{j=1}^{N} \left(
\gamma_{j,A,1} \gamma_{j,B,2} + \gamma_{j,A,2} \gamma_{j,B,1} \right) \nonumber \\
&+& \frac{i\Delta}{2} (1-\eta) \sum_{j=1}^{N-1} \left(
\gamma_{j,B,1} \gamma_{j+1,A,2} + \gamma_{j,B,2} \gamma_{j+1,A,1} \right)
\nonumber \\
& &
\eea

Taking from now on $\mu=0$ we have four special points, three
corresponding to topological phases and one respecting to a trivial
phase (Fig. \ref{fig1}):
i) Taking $\eta=-1$ and
$\Delta=0$ we have a state similar to the SSH or Schockley models with two
fermionic-like zero energy edge states, since the four operators
$\gamma_{1,A,1}, \gamma_{1,A,2}; \gamma_{N,B,1}, \gamma_{N,B,2}$ are missing from
the Hamiltonian.
ii) and iii) $\eta=0$ and $t=\pm \Delta$ are Kitaev like states since there are two Majorana operators
missing from the Hamiltonian, such as $\gamma_{1,A,1}$ and $\gamma_{N,B,2}$, one from each end.
iv) An example of a trivial phase is the point $\eta=1$ and $\Delta=0$ in which case there
are no zero energy edge states.
This model provides a testing ground for the comparison of fermionic and Majorana
edge modes.

The order parameters (also called topological correlators in Ref. \cite{bermudez})
can be determined separately for each phase.

\subsection{$\eta=-1, \Delta=0$ topological phase}

At the point $\mu=0,\eta=-1, \Delta=0$ shown in
Fig. \ref{fig1}, the Hamiltonian reduces to
\be H = it \sum_{j=1}^{N-1}
\left( \gamma_{j,B,2} \gamma_{j+1,A,1} - \gamma_{j,B,1}
\gamma_{j+1,A,2} \right).
\ee
Let us define non-local fermionic
operators~\cite{kitaev}
\bea
d_j &=& \frac{1}{2} \left(
\gamma_{j,B,2} + i \gamma_{j+1,A,1} \right),
\nonumber \\
d_j^{\dagger} &=& \frac{1}{2} \left( \gamma_{j,B,2} - i \gamma_{j+1,A,1} \right),
\label{eqkitaev}
\eea
and
\bea
f_j &=& \frac{1}{2} \left( \gamma_{j,B,1} - i \gamma_{j+1,A,2} \right),
\nonumber \\
f_j^{\dagger} &=& \frac{1}{2} \left( \gamma_{j,B,1} + i \gamma_{j+1,A,2} \right).
\eea
We can show that
\bea
i \gamma_{j,B,2} \gamma_{j+1,A,1} &=& 2 d_j^{\dagger} d_j -1, \nonumber \\
-i \gamma_{j,B,1} \gamma_{j+1,A,2} &=& 2 f_j^{\dagger} f_j -1.
\eea
In terms of these new operators we can write that
\be
H=t \sum_{j=1}^{N-1} \left( 2 d_j^{\dagger} d_j -1 +2 f_j^{\dagger} f_j -1 \right)
\ee
and, therefore, the Hamiltonian is diagonalized.
It is now clear that the ground state is obtained by taking
$d_j^{\dagger} d_j=0$ and $f_j^{\dagger} f_j=0$  at each site.
This new Hamiltonian in terms of the $d$ and $f$ operators is like
a Hamiltonian with no hopping and just a chemical potential
$\tilde{\mu}=-2t$.

The new operators can be related to the original ones in
terms of a non-local transformation as
\bea
d_j &=& \frac{i}{2}
\left( c_{j,B}^{\dagger} -c_{j,B} + c_{j+1,A} + c_{j+1,A}^{\dagger}
\right),\nonumber \\
f_j &=& \frac{1}{2} \left( c_{j,B}^{\dagger} +c_{j,B} - c_{j+1,A} + c_{j+1,A}^{\dagger} \right).
\label{eq:df}
\eea
Also
\bea
c_{j,A} &=& \frac{1}{2} \left[ -i(-d_{j-1}^{\dagger} +d_{j-1})-(f_{j-1} - f_{j-1}^{\dagger}) \right],
\nonumber \\
c_{j,B} &=& \frac{1}{2} \left[ f_{j}^{\dagger} +f_{j} +i( d_{j} + d_{j}^{\dagger}) \right].
\label{eq:cdf}
\eea
Note that the index $j$ of the $d$ and $f$ operators refers to
the bond connecting the $j,B$ and $j+1,A$ sites in the original representation.
At the special point we are considering we may also write
\be
H = -2t \sum_j \left( c_{j+1,A}^{\dagger} c_{j,B} + c_{j,B}^{\dagger} c_{j+1,A} \right).
\ee

In the diagonal basis the order parameter is
\bea
O_-^{SSH} &=& |0 0 \rangle \langle 0 0|, \nonumber \\
&=& I - |1 0\rangle \langle 1 0| - |0 1\rangle \langle 0 1| - |1 1\rangle \langle 1 1|, \nonumber \\
&=& 1-f^{\dagger}_j f_j \left( 1 -d_j^{\dagger} d_j \right) \nonumber \\
&-&d^{\dagger}_j d_j \left(1-f_j^{\dagger} f_j \right)  -f^{\dagger}_j f_j d^{\dagger}_j d_j, \nonumber \\
&=& \left( 1-f_j^{\dagger} f_j \right) \left( 1 -d_j^{\dagger} d_j \right).
\eea
These expressions are local in space. We may now use the relation
between the $d$ and $f$ operators and the original operators in Eq. (\ref{eq:df}). This
is a non-local transformation since it couples site $j$ with the nearest-neighbor
site $j+1$. The operator may now be obtained as
\bea
O_-^{SSH} &=& \frac{1}{2} \left( c_{j+1,A}^{\dagger} c_{j,B} + c_{j,B}^{\dagger} c_{j+1,A} \right)
\nonumber \\
&-& n_{j,B} n_{j+1,A} + \frac{1}{2} \left( n_{j,B} + n_{j+1,A} \right).
\label{eq:On}
\eea

\subsection{$\eta=0, \Delta=t$ topological phase}

Consider now the special point of the SSH-Kitaev model given by
$\eta=0, \Delta=t$.
This point is deep inside the Kitaev-like phase, as shown in Fig. \ref{fig1}.
At this point the Hamiltonian simplifies to
\be
H = t \sum_j \left( i \gamma_{j,A,2} \gamma_{j,B,1} +
i \gamma_{j,B,2} \gamma_{j+1,A,1} \right).
\ee
Write
\bea
i \gamma_{j,B,2} \gamma_{j+1,A,1} &=& 2 d_j^{\dagger}d_j-1,
\nonumber \\
i \gamma_{j,A,2} \gamma_{j,B,1} &=& 2 g_j^{\dagger}g_j-1.
\eea
Here the operator $d_j$ is defined in Eq. \ref{eqkitaev} and the operator $g_j$ is given by
\be
g_j = \frac{1}{2} \left( \gamma_{j,B,1} - i \gamma_{j,A,2} \right).
\ee
Therefore
\be
g_j = \frac{1}{2} \left( c_{j,B}^{\dagger} + c_{j,B} -c_{j,A}+c_{j,A}^{\dagger} \right).
\ee

The groundstate is obtained by taking $n_{d,j}=0$ and
$n_{g,j}=0$. Therefore let us define the new operator in the basis of these
occupation numbers  as
\bea
O_+^{SK} &=& |00\rangle \langle 00|,
\nonumber \\
&=& 1 -d_j^{\dagger} d_j - g_j^{\dagger} g_j +d_j^{\dagger} d_j g_j^{\dagger} g_j.
\eea
Using their expressions in terms of the original operators one obtains
\bea
O_+^{SK} &=& \frac{1}{4} \left[
c_{j,B}^{\dagger} \left( c_{j,A}+c_{j+1,A} \right)+
\left( c_{j,A}^{\dagger}+c_{j+1,A}^{\dagger} \right) c_{j,B} \right. \nonumber \\
&+& \left.
c_{j,B} \left( c_{j,A}-c_{j+1,A} \right)+
\left( c_{j,A}^{\dagger}-c_{j+1,A}^{\dagger} \right) c_{j,B}^{\dagger}
\right]
\nonumber \\
&-& \frac{1}{4} \left(2 n_{j,B}-1 \right) \left(
c_{j+1,A}^{\dagger} c_{j,A} + c_{j,A}^{\dagger} c_{j+1,A} \right. \nonumber \\
&+& \left. c_{j+1,A} c_{j,A} + c_{j,A}^{\dagger} c_{j+1,A}^{\dagger} \right)+\frac{1}{4}.
\eea

\subsection{$\eta=0,\Delta=-t$ topological phase}

Taking $\eta=0, \Delta=-t$, the Hamiltonian
reduces to
\be
H = -it \sum_j \left( \gamma_{j,A,1} \gamma_{j,B,2} + \gamma_{j,B,1} \gamma_{j+1,A,2} \right).
\ee

Define two new operators
\bea
f_j &=& \frac{1}{2} \left(\gamma_{j,B,1} -i \gamma_{j+1,A,2} \right),\\ 
h_j &=& \frac{1}{2} \left(\gamma_{j,A,1} -i \gamma_{j,B,2} \right). \\
\eea
In terms of the original fermion operators,
\bea
f_j &=& \frac{1}{2} \left[ c_{j,B} + c_{j,B}^{\dagger} -\left( c_{j+1,A} -c_{j+1,A}^{\dagger} \right) \right],
\nonumber \\
h_j &=& \frac{1}{2} \left[ c_{j,A} + c_{j,A}^{\dagger} -\left( c_{j,B} -c_{j,B}^{\dagger} \right) \right].
\eea

The Hamiltonian at this point can be written as
\be
H = t \sum_j \left( 2 h_j^{\dagger} h_j -1 +2 f_j^{\dagger} f_j -1 \right).
\ee
Again the groundstate is obtained by taking the state of no occupation of number operators of the $f$ and $h$ operators.
So let us define a new operator valid for negative $\Delta$ as
\bea
O_-^{SK} &=& |00\rangle \langle 00|,
\nonumber \\
&=& 1 -f_j^{\dagger} f_j - h_j^{\dagger} h_j +f_j^{\dagger} f_j h_j^{\dagger} h_j,\\
&=& \frac{1}{4} \left[
c_{j,B}^{\dagger} \left( c_{j,A}+c_{j+1,A} \right)+
\left( c_{j,A}^{\dagger}+c_{j+1,A}^{\dagger} \right) c_{j,B}
\right. \nonumber \\
&+& \left.
c_{j,B} \left( -c_{j,A}+c_{j+1,A} \right)+
\left( -c_{j,A}^{\dagger}+c_{j+1,A}^{\dagger} \right) c_{j,B}^{\dagger}
\right]
\nonumber \\
&-& \frac{1}{4} \left(2 n_{j,B}-1 \right) \left(
c_{j+1,A}^{\dagger} c_{j,A} + c_{j,A}^{\dagger} c_{j+1,A} \right. \nonumber \\
&-& \left. c_{j+1,A} c_{j,A} - c_{j,A}^{\dagger} c_{j+1,A}^{\dagger} \right)+\frac{1}{4}.
\eea

\subsection{$\eta=1, \Delta=0$ trivial phase}

For $\eta>0$, the mutual information is exponentially vanishing and
the correlation is not captured by considering the single-site block
with atoms A and B. However, one could take the block consisting of an
atom B at site $j$ and an atom A at site $j+1$. The mutual information
obtained would be the mirror image of that in Fig. 4 of reference \cite{us} about $\eta=0$.
The eigenspectrum of the reduced density matrix in this case is shown in Fig. 5(b) of the same reference. Carrying out
similar analysis as above, the order parameter takes the form of
Eq. (\ref{eq:On}), but with the index $\{j+1,A\}$ and $\{j,B\}$ being
replaced by $\{j,B\}$ and $\{j,A\}$, respectively. We have
\bea
O_+^{SSH} &=& \frac{1}{2} \left( c_{j,B}^{\dagger} c_{j,A} + c_{j,A}^{\dagger} c_{j,B} \right)
\nonumber \\
&-& n_{j,A} n_{j,B} + \frac{1}{2} \left( n_{j,A} + n_{j,B} \right).
\label{eq:Op}
\eea

\begin{figure*}
\centering
\includegraphics[width=0.95\linewidth]{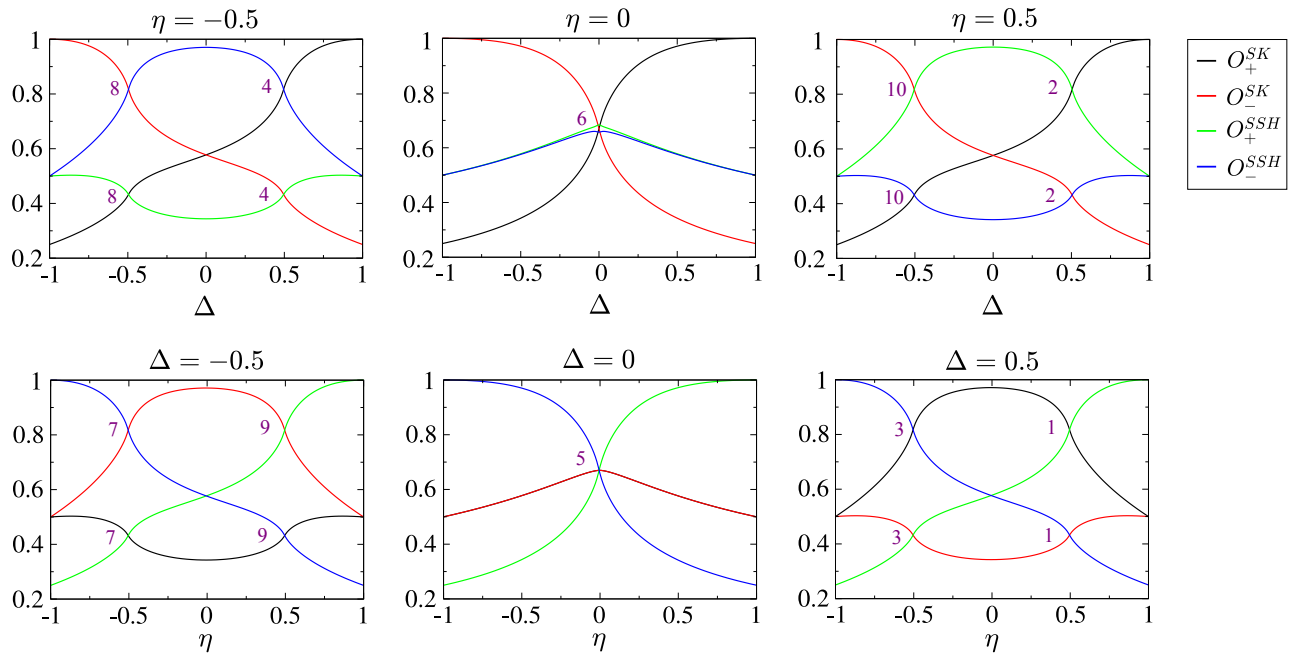}
\caption{Top panel: Order parameters as a function of $\Delta$ for constant values of $\eta=-0.5,0,0.5$. Bottom panel: Order parameters as a function of $\eta$ for constant values of $\Delta=-0.5,0,0.5$. We consider open boundary conditions and a system size $N=100$. The numbers in purple next to where the order parameters cross correspond to the cuts through the phase boundaries label in Fig. \ref{fig1}(a) respectively.}
\label{fig:fig2}
\end{figure*}

%
%

\section{Phase diagram and order parameters}
\label{sec:orderparameter}

\subsection{Calculation method}

We may now calculate the order parameters (or topological correlators)
as the groundstate average values of the operators $O_+^{SSH}, O_-^{SSH},
O_+^{SK}, O_-^{SK}$ defined in the previous section.

The average values may be obtained for instance using exact diagonalization,
density matrix renormalization group \cite{DMRG1,DMRG2,DMRG3} (particularly useful if one introduces
interactions between the original fermions) or via the solution of
the Bogoliubov-de Gennes (BdG) equations in the absence of interactions.
This last method allows the solution for large systems (suited for the finite size scaling analysis carried out ahead).
We consider open boundary conditions in the followings. The results obtained using exact diagonalization or the BdG method agree.

We may write that
\bea
c_{j,A} &=& \sum_n \left( u_{j,A}^n \gamma_n + v_{j,A}^n \gamma_n^{\dagger} \right)
\nonumber \\
c_{j,B} &=& \sum_n \left( u_{j,B}^n \gamma_n + v_{j,B}^n \gamma_n^{\dagger} \right)
\eea
where $\gamma_n$ are the fermionic operators that diagonalize the Hamiltonian.
The Bogoliubov-de Gennes equations are written in real space as
\be
\sum_{j^{\prime}} H_{j,j^{\prime}}
\left( \begin{array}{c}
 u_{j^{\prime},A} \\  u_{j^{\prime},B} \\ v_{j^{\prime},A} \\  v_{j^{\prime},B}
\end{array} \right) =
\epsilon_n
\left( \begin{array}{c}
 u_{j,A} \\  u_{j,B} \\ v_{j,A} \\  v_{j,B}
\end{array} \right)
\ee
where $\epsilon_n$ are the energy eigenvalues and $u_{A,B}$ and $v_{A,B}$ are the components
of the eigenfunctions and $j^{\prime}$ is restricted to $j=j^{\prime}$ and $j=j^{\prime} \pm 1$.

The averages of the order parameters may then be obtained by solving the BdG equations and determining
the wave functions. Considering a finite system of size $N$, the problem requires the diagonalization
of a $(4N)\times (4N)$ matrix.

\subsection{Results}

We may now consider cuts in the phase diagram and calculate the order parameters. With open boundary conditions it is better to take average over alternating lattice sites \cite{bermudez}.
Specifically
\be
O=\frac{2}{N} \sum_{j=2i+1} O_j.
\label{eq:avgO}
\ee

In Fig. \ref{fig:fig2} top panel we consider three cuts for $\eta=-0.5, \eta=0, \eta=0.5$ and in the bottom panel we consider cuts for $\Delta=-0.5, \Delta=0, \Delta=0.5$. We calculate the averaged order parameter and the results shown are for a large system size of $N=100$. The results for the two types of cuts are quite symmetrical if we change $\eta$ to $\Delta$ and vice-versa changing also appropriately the order parameters. The order parameters clearly identify various phases in the model and we observe the follows.
\begin{enumerate}[i]
	\item At the points where each order parameter is defined the order parameter is normalized to one since it is the groundstate average value of the projector to that state.
	\item At each phase the order parameter characteristic of that phase has the largest value.
	\item The order parameters cross at the transition lines.
\end{enumerate}

\begin{figure*}[t]
	\includegraphics[width=0.8\textwidth]{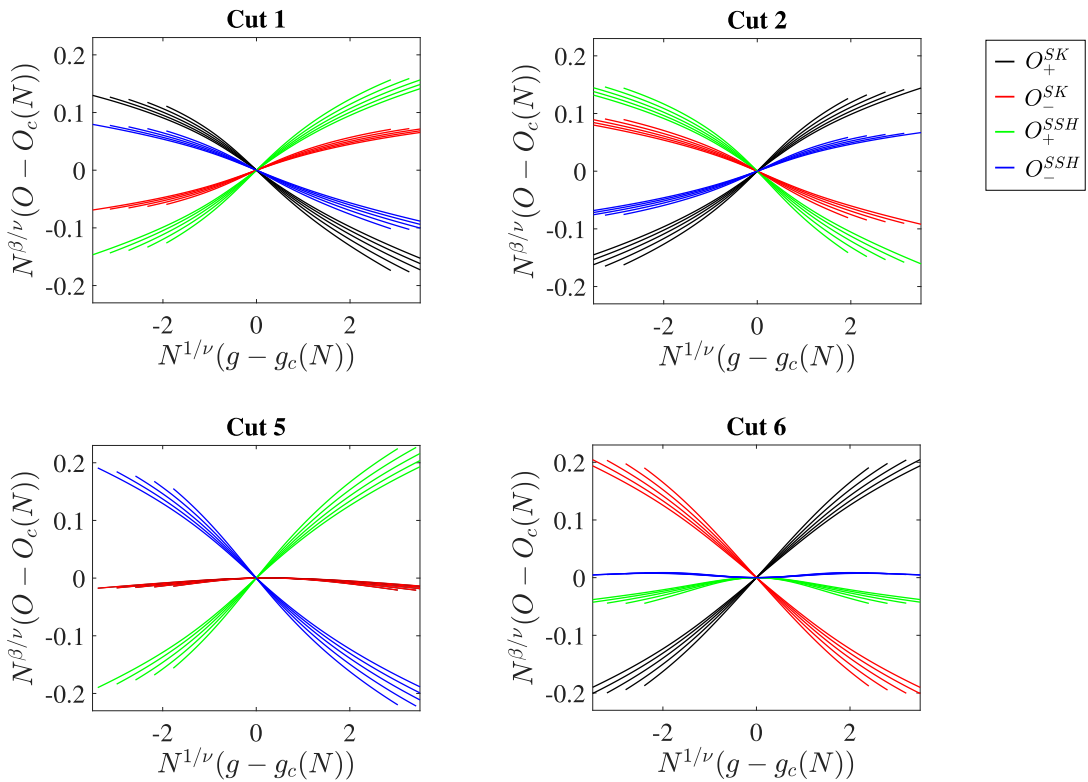}
	\caption{\label{fig5}
		(Color online)
		Scaling of various order parameters for cuts 1, 2, 5, 6 with $\nu=1, \beta=1/8$. The black, red, green, and blue curve corresponds to $O_+^{SK}, O_-^{SK}, O_+^{SSH}$, and $O_-^{SSH}$, respectively. Here we use OBC and order parameters are averaged over all odd sites (or even sites) as defined in Eq. \ref{eq:avgO}. We consider system sizes $N=24,28,32,36,40$.
	}
\end{figure*}

\section{Scaling and critical exponents}
\label{sec:scaling}

We may now determine the scaling properties of the order parameters.
The scaling allows to determine the critical exponents and the universality
class of the system.

We recall that the order parameters defined above do not vanish at the transition
points as usual. Also the order parameter does not separate a disordered phase
with a vanishing order parameter from an ordered phase with finite values of
the order parameter. The various averages of the order operators that represent the various topological phases cross at the driving parameter $g=g_c(N)$ with finite values $O_c(N)$.
The scaling is then expected to be of the form
\be
N^{\beta/\nu} (O-O_c) = f \left( N^{1/\nu} (g-g_c) \right).
\label{eq:scaling}
\ee
Here $\beta$ is the usual critical exponent associated with an order parameter and $\nu$ is the exponent associated with the correlation
length. The function $f$ is the
scaling function. Plotting the left hand side of the scaling relation
against the argument of the scaling function, we expect that the curves
for different system sizes should collapse into a single curve near the
critical point. Note that the scaling function is independent of the system size
at the crossing point $g=g_c$, and the size dependence of each order parameter
is such that in the thermodynamic limit the order parameter should converge
to a value $O_c\neq 0$ in our problem.
%

Consider first a single-band Kitaev model.
The model is not expected to have some form of quasi-long range
order. The correlation functions decay exponentially with a correlation
length that indeed diverges at the topological transition with an
exponent $\nu=1$. This may be obtained using the scaling behavior of the energy gap $E_g\sim (g-g_c)^{\nu z}$, where $z$ is the dynamical critical
exponent. Since the energy spectrum is linear, we have $z=1$ and the gap scales linearly which leads to $\nu=1$ (as shown for instance in \cite{continentino1}).
Generalizing the Kitaev model to a multi-band model with an anti-symmetric
coupling between the two bands leads to a rich phase diagram that displays
a topological transition between a Weyl-like phase and a conventional
superconductor that turns out to be in a different universality class of the
Kitaev model \cite{continentino2}. The dispersion relation near the transition points turns out
to be quadratic leading to a dynamical critical exponent $z=2$ and since
the gap as a function of the driving parameter (the chemical potential)
vanishes linearly, this leads to $\nu=1/2$ (also using the hyperscaling
relation $2-\alpha=\nu (d+z)$, where $d$ is the spatial dimension leads to
$\alpha=3/2$, while in the Kitaev universality class $\alpha=0$).

Consider now the SSH model with no superconductivity.
In ref. \cite{bermudez} a mapping was established in some regime between
the Schwinger model on a lattice and the SSH model. Using the order parameter
$O_-^{SSH}$ it was shown that the model
is in the universality class of the $d=2$ classical
Ising model or the quantum Ising model in a transverse field
(recall that a mapping exists between a quantum model in $d$
dimensions and a classical model in $d+z$ dimensions; therefore if $z=1$ there
is a mapping from a quantum one-dimensional model and a classical two-dimensional
model). Note that the two-dimensional Ising model displays a true phase transition
and it makes sense to define an order parameter. In this class the critical exponents
are given by $\nu=1, \beta=1/8$.
The results are consistent with previous treatments of the massive
Schwinger model \cite{ising1,ising2,ising3}.

\subsection{Scaling in the Ising universality class}

Let us now consider the SSH model with triplet pairing and consider open
boundary conditions (ignoring the small discontinuity at the transition
points observed for small system sizes yields similar results in the case of
periodic boundary conditions). In order to consider the scaling we must choose the critical
exponents and perform the analysis with different system sizes.

Consider some cuts in the phase diagram as indicated by the purple dash segments in Fig. \ref{fig1}(a).
In Fig. \ref{fig5}, we show results for the scaling of the order parameters for some of the
various cuts considered.
Here we use the Ising universality class with $\nu=1$ and $\beta=1/8$ as
obtained before for the SSH model.
The scaling seems to work approximately well for the various cases since the curves near the transition points collapse basically into a single curve.

\subsection{Optimizing the scaling}

\begin{figure}[t]
\includegraphics[width=0.9\columnwidth]{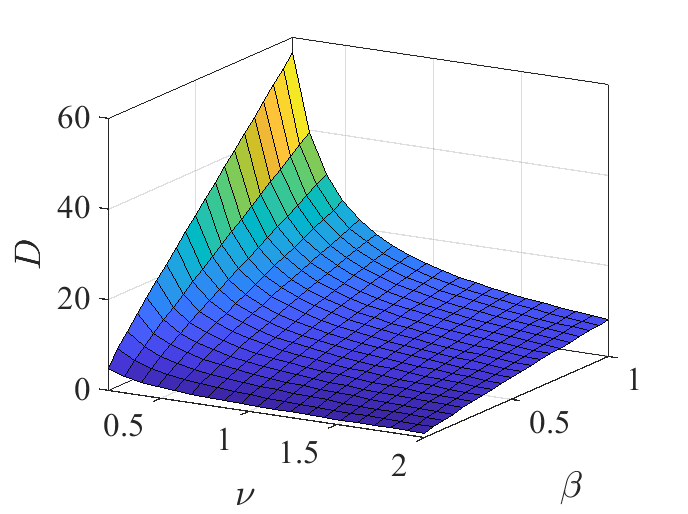}
\caption{\label{fig6}
(Color online)
Minimization of spreading of scaled curves of $O_+^{SK}$ around the critical point of cut 1.
}
\end{figure}

While the choice of critical exponents above describes well the scaling of the various
order parameters near the various transition lines,
we may find a pair of critical exponents that best fit the scaling ansatz in Eq. \ref{eq:scaling}.
A criteria may be used such that the deviations between the various curves for
different system sizes in the vicinity of the critical point is minimized. One may also minimize the squares of these deviations.
The logarithm of this
deviation $D$ is shown in Fig. \ref{fig6}, for the scaling of the order
parameter $O_+^{SK}$ across the transition associated with cut 1.
Specifically for each value of the driving parameter around the critical point (in this case $\eta$), the squared differences between the minimum and maximum value of the
order parameter for the various system sizes (taken here as $N=24,40,60,100,200$) are considered.
The result shown in Fig. \ref{fig6} is the logarithm of the sum of this differences squared at each $\eta$ value.
The results obtained suggest that the universality class may be different from the Ising
class.
We may also fit a polynomial
function to the results for the various system sizes and use the least squares
method for a given scaling function. This leads to similar results.
Similar results are also found for the other cuts considered in Fig. \ref{fig1}(a), and for the
various order parameters
and therefore are not shown here.

As we can see the deviation is minimized if we keep increasing $\nu$ and decreasing $\beta$.
So a value of $\nu=1$ is a large value and $\beta=1/8$ is small but the agreement becomes
better (but slowly varying) if we change along the lines mentioned. Note that $\beta=0$ and a very large $\nu$ leads to a scaling of the type
\be
O-O_c = f \left( (g-g_c) \right)
\ee
with no system size dependence at all. Clearly in the infinite system limit
this holds.

The exponent $\nu=1$ has
also been obtained by other methods. In one dimensional systems of the Dirac type of
class $AIII$ it has been shown \cite{sigrist} that in general $\gamma=\nu$. Considering the
case of the SSH model it was explicitly shown \cite{sigrist} that $\nu=1$.
Our results for cut 5 (with $\Delta=0$ and changing $\eta$ as in the SSH model)
as for the other cuts are not inconsistent with the analytical behavior for the
SSH model since the optimization shows a very slow change of the least-squares deviation.

\section{Local Hamiltonian as "order parameter" \label{seclh}}
\label{sec:localH}

\subsection{Single band Kitaev model}

Consider the single-band Kitaev model at a zero chemical potential described by the Hamiltonian
\be
H=-t \sum_j \left (c_{j+1}^{\dagger} c_j + c_j^{\dagger} c_{j+1} \right)
+ \Delta \sum_j \left( c_j c_{j+1} +c_{j+1}^{\dagger} c_j^{\dagger} \right).
\ee
The Hamiltonian takes a simple form at the points $\Delta=t$ and
$\Delta=-t$, as discussed above for the generalization of the model
to two sublattices.
A similar procedure allows to determine two operators associated with
the diagonalization of the Hamiltonian at these two points.
The order parameters are simply given by the average of the projectors to a single-site
zero occupation of the $d_j$ operators defined in eq. \ref{eqkitaev} (and the
corresponding for the case of $\Delta=-t$). We obtain then that
\bea
O_{j,+}^{K} &=& |0\rangle \langle 0| = 1- d_j^{\dagger} d_j \nonumber \\
&=& \frac{1}{2} -\frac{1}{2} H_j(t=\Delta=1)
\eea
and
\be
O_{j,-}^{K} =
\frac{1}{2} -\frac{1}{2} H_j(t=-\Delta=1)
\ee
where $H_j$ is the contribution from site $j$ to the Hamiltonian
\be
H_j(t=\Delta=1) = -c_{j+1}^{\dagger} c_j - c_j^{\dagger} c_{j+1}
+ c_j c_{j+1} +c_{j+1}^{\dagger} c_j^{\dagger}
\ee
and
\be
H_j(t=-\Delta=1) = -c_{j+1}^{\dagger} c_j - c_j^{\dagger} c_{j+1}
- c_j c_{j+1} -c_{j+1}^{\dagger} c_j^{\dagger},
\ee
respectively. Diagonalizing the full Hamiltonian at an arbitrary
point in the phase diagram we may write that
\bea
O_{j,+}^{K} &=& \frac{1}{2}+\sum_n v_j^n \left( u_{j+1}^n + v_{j+1}^n \right),
\nonumber \\
O_{j,-}^{K} &=& \frac{1}{2}+\sum_n v_j^n \left( -u_{j+1}^n + v_{j+1}^n \right).
\eea

\begin{figure}
\includegraphics[width=0.75\columnwidth]{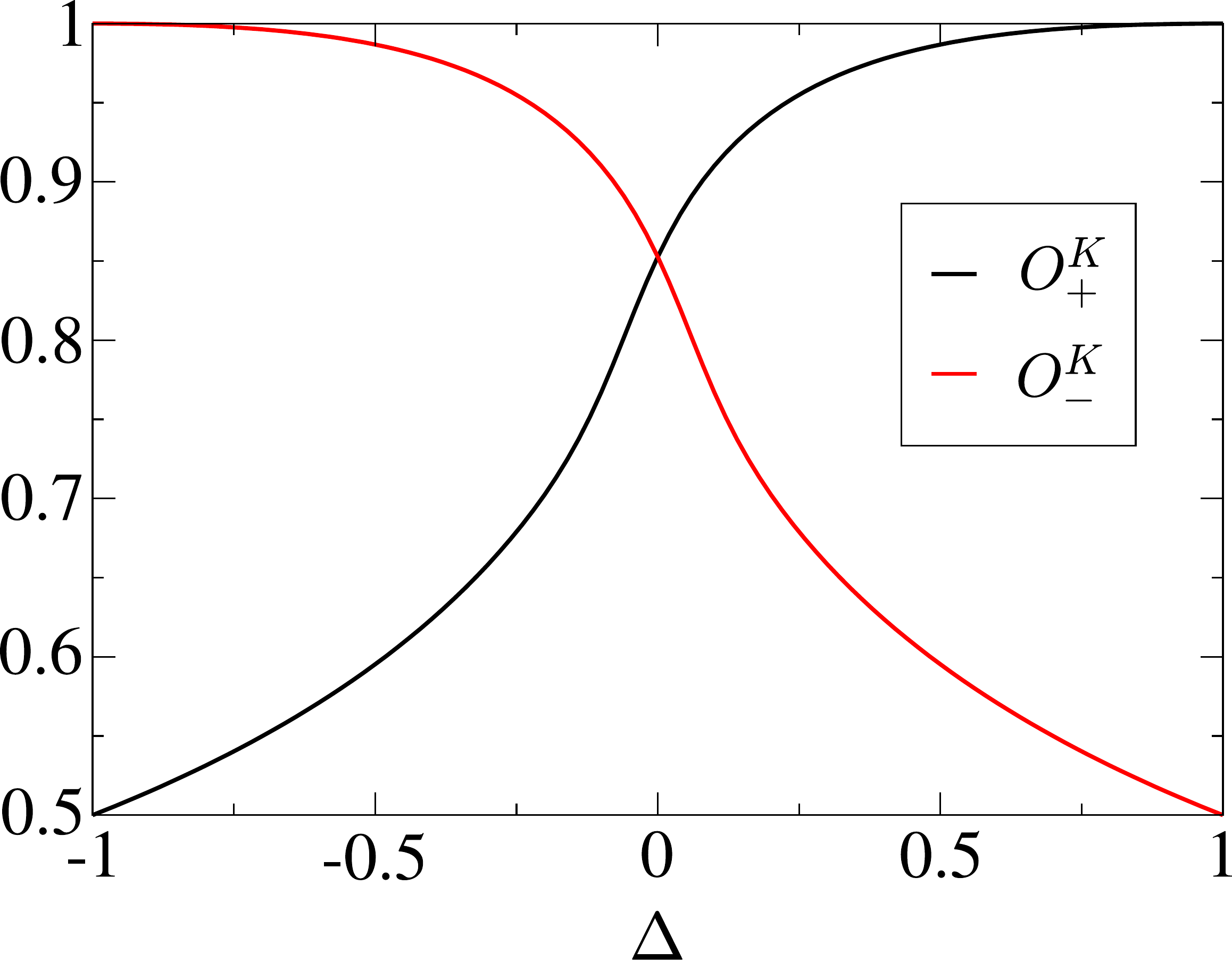}
\caption{\label{figenerg1}
(Color online)
Order parameters for the single band Kitaev model as a function
of $\Delta$ for $N=32$.
}
\end{figure}

In Fig. \ref{figenerg1} we calculate the order parameters as a function of $\Delta$ and take the average over odd sites as in the previous section. The order parameters cross at the transition point. Since these operators are basically the local Hamiltonian
plus a constant this suggests that the local Hamiltonian itself may be used as
an order parameter. While in the case of the single-band Kitaev model
the procedure to determine the projectors leads to the local Hamiltonian
at that specific point in the phase diagram, this does not occur in the
SSH-Kitaev model, as shown in Sec. \ref{sec:model}.

In any case, let us consider the local Hamiltonian of the SSH-Kitaev model
to see if it can be used as a suitable operator that leads to an order
parameter.

\subsection{SSH-Kitaev model}

\begin{figure}
\includegraphics[width=0.8\columnwidth]{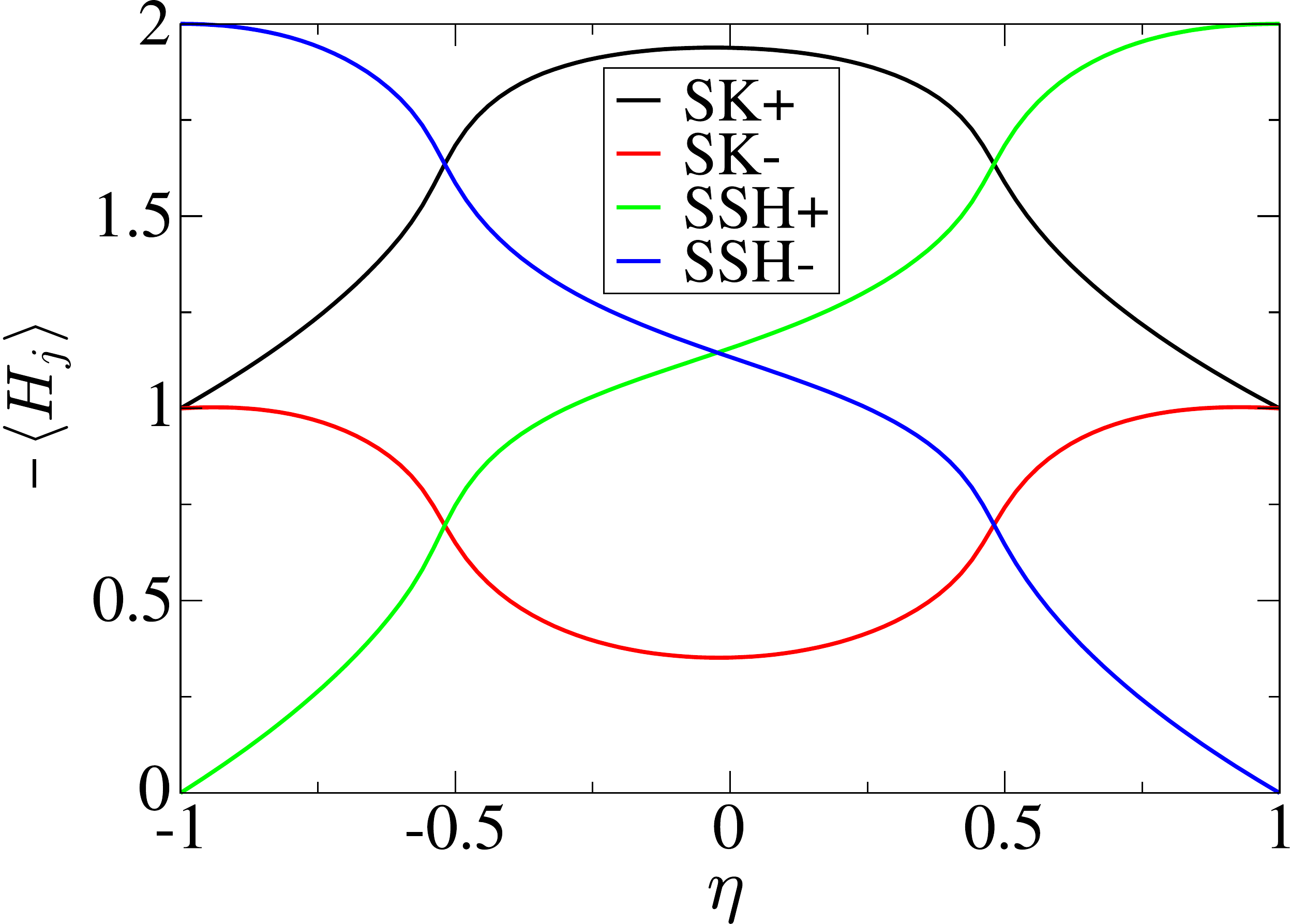}
\caption{\label{figenergy2}
(Color online)
Local Hamiltonian order parameters for the SSH-Kitaev model as a function
of $\eta$ at $\Delta=0.5$, using OBC and taking $N=32$.
}
\end{figure}

Write the Hamiltonian in Eq. \ref{ham1} as $H=\sum_j H_j$ and then
define four operators as the local Hamiltonian, $H_j$, at the points indicated
in Fig. \ref{fig1}(a).
We calculate their averages at an arbitrary point of the phase diagram using the
eigenstates of the Hamiltonian at this arbitrary point. Note that in general
these states are not the eigenstates of the Hamiltonians at the special points
marked in Fig. \ref{fig1}(a). As an example let us consider that we fix $\Delta=0.5$
and change $\eta$ from $-1$ to $1$. The results for other ranges of values lead
to similar conclusions.
In Fig. \ref{figenergy2} we show the results for this cut in the phase diagram for
the four local Hamiltonians. The results are strikingly similar to the ones obtained
using the order parameters $O_+^{SK}, O_-^{SK}, O_+^{SSH}, O_-^{SSH}$ and the transitions
are clearly signaled by the crossings of the various order parameters defined from the local
Hamiltonians.

We may as well identify the phase transitions considering only one of the
order parameters and actually we do not have to limit their definition
to the special points where a diagonalization of the Hamiltonian can be
performed analytically (as easily done using the Majorana representation).
Let us consider the local Hamiltonian, $H_j$, in an arbitrary point in the phase
diagram of the SSH-Kitaev model (this should hold for an arbitrary
Hamiltonian). Let us now consider a cut in the phase diagram that crosses some
transition or transition lines (points). We will show now that the derivative of the average of
this local Hamiltonian in the basis of the eigenstates of the full Hamiltonian
at each point along the cut with respect to the parameter that defines the cut,
detects the transition lines.
We will focus on some example but the result can be checked for arbitrary examples.

\begin{figure}[t]
\includegraphics[width=0.8\columnwidth]{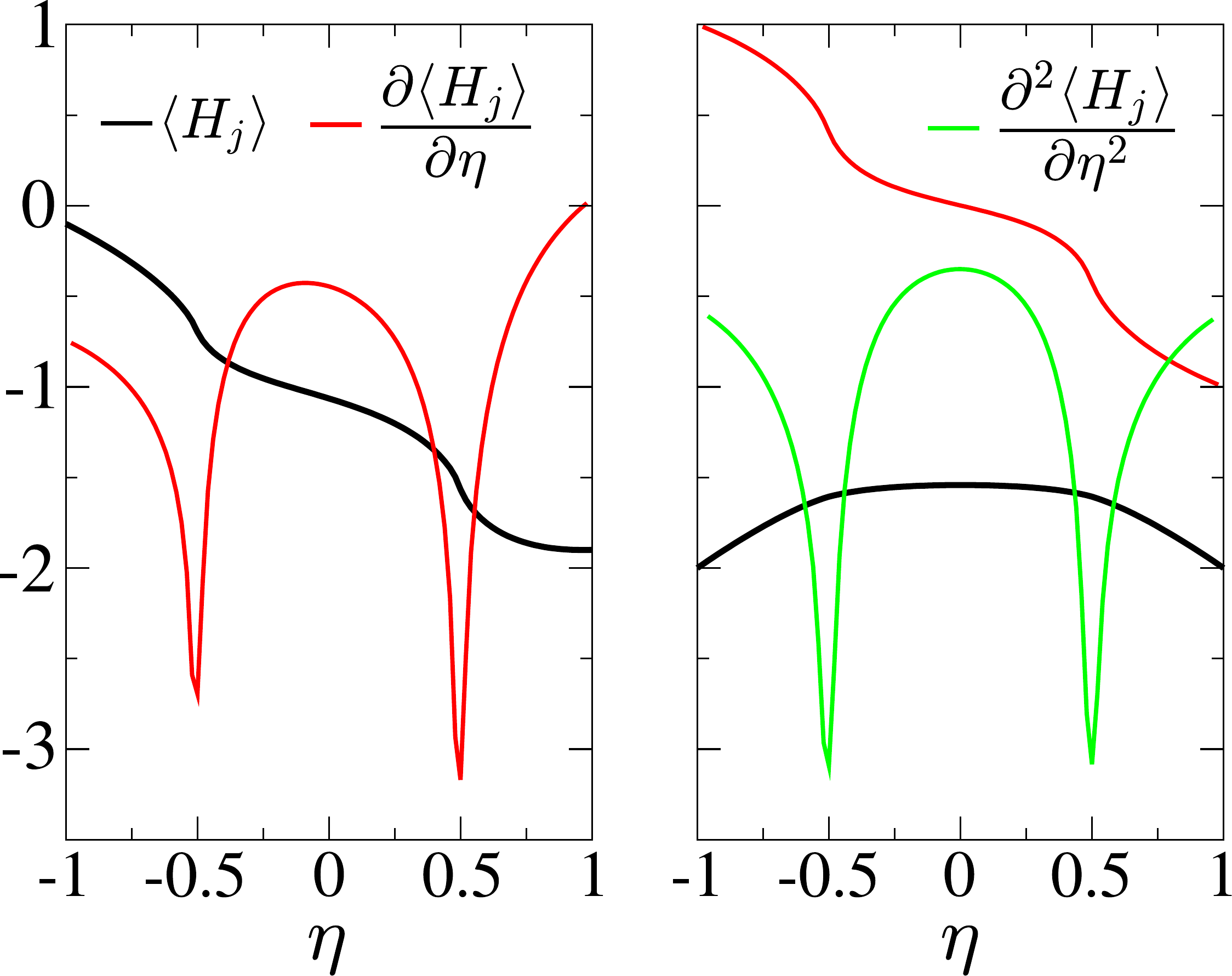}
\caption{\label{figenerg3}
(Color online)
Local Hamiltonian order parameters for the SSH-Kitaev model and its
derivatives with respect to $\eta$ as a function
of $\eta$ at $\Delta=0.5$, using OBC and taking $N=100$. In the left panel, the local Hamiltonian at the point $\eta=0.9$ and $\Delta=-0.1$ is considered. In the right panel
we take the local Hamiltonian at $\Delta=0.5$ and at the sequence of values of $\eta$
along the cut. Therefore, in the right panel it is the energy per site.
}
\end{figure}

Consider for example the same cut as above, where $\Delta=0.5$ and we change
$\eta$ from $-1$ to $1$. There are two transition points at $\eta=-0.5$ and $\eta=0.5$.
Consider the local Hamiltonian at the point $\eta=0.9, \Delta=-0.1$, some arbitrary point
in the phase diagram and not on the cut that we choose.
We also calculate the average energy per site. The results are shown in Fig. \ref{figenerg3}.
While the transitions are detected by calculating the second derivative of the energy per site
with respect to the driving parameter, $\eta$, (this is like a susceptibility or
related to the fidelity susceptibility) it is enough to calculate the first derivative
of the average of the local Hamiltonian at an arbitrary point in the phase diagram
not necessarily located in a point on the cut.

\section{Conclusions}
\label{sec:conlusion}

With the method introduced, we obtained the order parameters that clearly signal the various phase transitions in
the SSH model with triplet pairing. Also the magnitudes of the various order parameters
are in complete agreement with the sequence of phases in the sense that the larger
order parameter corresponds to the dominant characteristic of each phase. We expect our method can also be applied to other models, for example the multiband hybridized superconductors \cite{continentino2,continentino3}, that has a diagonal representation of the Hamiltonian at specific points of the phase diagram.

The finite-size scaling results with exponents $\nu=1$ and $\beta=1/8$ seem to support that the model belongs to the quantum Ising
universality class. This is consistent
with previous results obtained for the SSH model. However, the least square optimization analysis also showed that $\nu$ and $\beta$ can be values larger than $1$ and $1/8$ respectively. This suggests some non-trivial scaling relations may be required to describe the quantum criticality in the model since we have competing order parameters.
One possibility is related to the existence of more than one order parameter
as discussed for instance in reference \cite*{chaikin}. Another possibility is the existence of more than one correlation length as discussed in reference \cite{shao}.

\section*{ACKNOWLEDGEMENTS}
We acknowledge support from the National Research Foundation and the Ministry of Education of Singapore, NSAF U1530401 and computational resources from the Beijing Computational Science Research Center. This research was also partially funded by Polisimulator project co-financed by Greece and the EU Regional Development Fund. PDS acknowledges partial support from FCT through grant UID/CTM/04540/2013.

\end{document}